%%%%%%%%%%%%%%%%%%%%%%%%%%%%%%%%%%%%%%%%%%%%%%%%%%%%%%%%%%%%%%%%%
%                                                               %
%       FONT FAMILIES:                                          %
%                                                               %
%%%%%%%%%%%%%%%%%%%%%%%%%%%%%%%%%%%%%%%%%%%%%%%%%%%%%%%%%%%%%%%%%
%                                                               %
%       Define script letters as rsfs                           %
%               (or redefine as cal)                            %
%                                                               %
%                                                               %
%%%%%%%%%%%%%%%%%%%%%%%%%%%%%%%%%%%%%%%%%%%%%%%%%%%%%%%%%%%%%%%%%
\newfam\scrfam
\batchmode\font\tenscr=rsfs10 \errorstopmode
\ifx\tenscr\nullfont
        \message{rsfs script font not available. Replacing with calligraphic.}
        
\else   
        \font\sevenscr=rsfs7
        \font\fivescr=rsfs5
        \skewchar\tenscr='177 \skewchar\sevenscr='177 \skewchar\fivescr='177
        \textfont\scrfam=\tenscr \scriptfont\scrfam=\sevenscr
        \scriptscriptfont\scrfam=\fivescr

\fi
%%%%%%%%%%%%%%%%%%%%%%%%%%%%%%%%%%%%%%%%%%%%%%%%%%%%%%%%%%%%%%%%%
%                                                               %
%       Blackboard bold (or redefine as boldface)               %
%                                                               %
%%%%%%%%%%%%%%%%%%%%%%%%%%%%%%%%%%%%%%%%%%%%%%%%%%%%%%%%%%%%%%%%%
\newfam\msbfam
\batchmode\font\twelvemsb=msbm10 scaled\magstep1 \errorstopmode
\ifx\twelvemsb\nullfont\def\Bbb{\bf}
        
	\font\eightbbb=cmb10 at 8pt
	\message{Blackboard bold not available. Replacing with boldface.}
\else   \catcode`\@=11
        \font\tenmsb=msbm10 \font\sevenmsb=msbm7 \font\fivemsb=msbm5
        \textfont\msbfam=\tenmsb
        \scriptfont\msbfam=\sevenmsb \scriptscriptfont\msbfam=\fivemsb
        \def\Bbb{\relax\expandafter\Bbb@}
        \def\Bbb@#1{{\Bbb@@{#1}}}
        \def\Bbb@@#1{\fam\msbfam\relax#1}
        \catcode`\@=\active
	
	\font\eightbbb=msbm8
\fi
%%%%%%%%%%%%%%%%%%%%%%%%%%%%%%%%%%%%%%%%%%%%%%%%%%%%%%%%%%%%%%%%%
%                                                               %
%       MORE FONTS:                                             %
%                                                               %
%%%%%%%%%%%%%%%%%%%%%%%%%%%%%%%%%%%%%%%%%%%%%%%%%%%%%%%%%%%%%%%%%
        \font\eightrm=cmr8              \def\xrm{\eightrm}
        \font\eightbf=cmbx8             \def\xbf{\eightbf}
        \font\eightit=cmti10 at 8pt     \def\xit{\eightit}
%%%     \font\eightit=cmti8             \def\xit{\eightit}
        \font\eighttt=cmtt8             \def\xtt{\eighttt}
        \font\eightcp=cmcsc8
        \font\eighti=cmmi8              \def\xold{\eighti}
        \font\teni=cmmi10               \def\old{\teni}
        \font\tencp=cmcsc10
        \font\tentt=cmtt10
        \font\twelverm=cmr12
        \font\twelvecp=cmcsc10 scaled\magstep1
        \font\fourteencp=cmcsc10 scaled\magstep2

\def\noblackbox{\overfullrule=0pt}
\noblackbox

%%%%%%%%%%%%%%%%%%%%%%%%%%%%%%%%%%%%%%%%%%%%%%%%%%%%%%%%%%%%%%%%%
%                                                               %
%       HEADLINE:                                               %
%                                                               %
%%%%%%%%%%%%%%%%%%%%%%%%%%%%%%%%%%%%%%%%%%%%%%%%%%%%%%%%%%%%%%%%%
\newtoks\headtext
\headline={\ifnum\pageno=1\hfill\else
	\ifodd\pageno{\eightcp\the\headtext}\dotfill{ }{\old\folio}
	\else{\old\folio}{ }\dotfill{\eightcp\the\headtext}\fi
\fi}
\def\makeheadline{\vbox to 0pt{\vss\noindent\the\headline\break
\hbox to\hsize{\hfill}}
        \vskip2\baselineskip}
%%%%%%%%%%%%%%%%%%%%%%%%%%%%%%%%%%%%%%%%%%%%%%%%%%%%%%%%%%%%%%%%%
%                                                               %
%       FOOTNOTES:                                              %
%                                                               %
%%%%%%%%%%%%%%%%%%%%%%%%%%%%%%%%%%%%%%%%%%%%%%%%%%%%%%%%%%%%%%%%%
\newcount\infootnote
\infootnote=0
\def\foot#1#2{\infootnote=1
\footnote{$\,{}^{#1}$}{\vtop{\baselineskip=.75\baselineskip
\advance\hsize by -\parindent\noindent{\xrm #2}}}\infootnote=0}
%%%%%%%%%%%%%%%%%%%%%%%%%%%%%%%%%%%%%%%%%%%%%%%%%%%%%%%%%%%%%%%%%
%                                                               %
%       REFERENCES:                                             %
%                                                               %
%%%%%%%%%%%%%%%%%%%%%%%%%%%%%%%%%%%%%%%%%%%%%%%%%%%%%%%%%%%%%%%%%
\newcount\refcount
\refcount=1
\newwrite\refwrite
\def\oldsize{\ifnum\infootnote=1\xold\else\old\fi}
\def\ref#1#2{
	\def#1{{{\oldsize\the\refcount}}\ifnum\the\refcount=1\immediate\openout\refwrite=\jobname.refs\fi\immediate\write\refwrite{\item{[{\xold\the\refcount}]} 
	#2\hfill\par\vskip-2pt}\xdef#1{{\oldsize\the\refcount}}\global\advance\refcount by 1}
	}
\def\refout{\catcode`\@=11
        \xrm\immediate\closeout\refwrite
        \vskip2\baselineskip
        {\noindent\twelvecp References}\hfill\vskip\baselineskip
                                                %\vskip.25\baselineskip%%%%
        %\parskip=.875\parskip
        %\baselineskip=.8\baselineskip
        \baselineskip=.75\baselineskip
        \input\jobname.refs
        %\parskip=8\parskip \divide\parskip by 7
        %\baselineskip=1.25\baselineskip
        \baselineskip=4\baselineskip \divide\baselineskip by 3
        \catcode`\@=\active\rm}

\def\hepth#1{\href{http://xxx.lanl.gov/abs/hep-th/#1}{{\xtt hep-th/#1}}}
\def\jhep#1#2#3{\href{http://jhep.sissa.it/stdsearch?paper=#1\%28#2\%29#3}{{\xit JHEP} {\xbf #1} ({\xold#2}) {\xold#3}}}
\def\PLB#1#2#3{Phys. Lett. {\xbf B#1} ({\xold#2}) {\xold#3}}
\def\NPB#1#2#3{Nucl. Phys. {\xbf B#1} ({\xold#2}) {\xold#3}}
\def\PRD#1#2#3{Phys. Rev. {\xbf D#1} ({\xold#2}) {\xold#3}}
\def\PRL#1#2#3{Phys. Rev. Lett. {\xbf #1} ({\xold#2}) {\xold#3}}
\def\IJMP#1#2#3{Int. J. Mod. Phys. {\xbf #1} ({\xold#2}) {\xold#3}}

\def\JGP#1#2#3{J. Geom. Phys. {\xbf #1} ({\xold#2}) {\xold#3}}
%%%%%%%%%%%%%%%%%%%%%%%%%%%%%%%%%%%%%%%%%%%%%%%%%%%%%%%%%%%%%%%%%
%                                                               %
%       SECTION NUMBERING:                                      %
%                                                               %
%%%%%%%%%%%%%%%%%%%%%%%%%%%%%%%%%%%%%%%%%%%%%%%%%%%%%%%%%%%%%%%%%
\newcount\sectioncount
\sectioncount=0
\def\section#1#2{\global\eqcount=0
	\global\subsectioncount=0
        \global\advance\sectioncount by 1
        \xdef#1{{\old\the\sectioncount}}
	\vskip2\baselineskip\noindent
        \line{\twelvecp\the\sectioncount. #2\hfill}
	\vskip\baselineskip\noindent}
\newcount\subsectioncount
\def\subsection#1#2{\global\advance\subsectioncount by 1
	\xdef#1{{\old\the\sectioncount}.{\old\the\subsectioncount}}
	\vskip.8\baselineskip\noindent
	\line{\tencp\the\sectioncount.\the\subsectioncount. #2\hfill}
	\vskip.5\baselineskip\noindent}
\newcount\appendixcount
\appendixcount=0
\def\appendix#1{\global\eqcount=0
        \global\advance\appendixcount by 1
        \vskip2\baselineskip\noindent
        \ifnum\the\appendixcount=1
        \hbox{\twelvecp Appendix A: #1\hfill}\vskip\baselineskip\noindent\fi
    \ifnum\the\appendixcount=2
        \hbox{\twelvecp Appendix B: #1\hfill}\vskip\baselineskip\noindent\fi
    \ifnum\the\appendixcount=3
        \hbox{\twelvecp Appendix C: #1\hfill}\vskip\baselineskip\noindent\fi}

%%%%%%%%%%%%%%%%%%%%%%%%%%%%%%%%%%%%%%%%%%%%%%%%%%%%%%%%%%%%%%%%%
%                                                               %
%       EQUATION NUMBERING                                      %
%                                                               %
%%%%%%%%%%%%%%%%%%%%%%%%%%%%%%%%%%%%%%%%%%%%%%%%%%%%%%%%%%%%%%%%%
\newcount\eqcount
\eqcount=0
\def\Eqn#1{\global\advance\eqcount by 1
        \xdef#1{{\oldsize\the\sectioncount}.{\oldsize\the\eqcount}}
        \ifnum\the\appendixcount=0
                \eqno({\oldstyle\the\sectioncount}.{\oldstyle\the\eqcount})\fi
        \ifnum\the\appendixcount=1
                \eqno({\oldstyle A}.{\oldstyle\the\eqcount})\fi
        \ifnum\the\appendixcount=2
                \eqno({\oldstyle B}.{\oldstyle\the\eqcount})\fi
        \ifnum\the\appendixcount=3
                \eqno({\oldstyle C}.{\oldstyle\the\eqcount})\fi}
\def\eqn{\global\advance\eqcount by 1
        \ifnum\the\appendixcount=0
                \eqno({\oldstyle\the\sectioncount}.{\oldstyle\the\eqcount})\fi
        \ifnum\the\appendixcount=1
                \eqno({\oldstyle A}.{\oldstyle\the\eqcount})\fi
        \ifnum\the\appendixcount=2
                \eqno({\oldstyle B}.{\oldstyle\the\eqcount})\fi
        \ifnum\the\appendixcount=3
                \eqno({\oldstyle C}.{\oldstyle\the\eqcount})\fi}
\def\multi{\global\advance\eqcount by 1}
\def\multieq#1#2{\xdef#1{{\oldsize\the\eqcount#2}}
        \eqno{({\oldstyle\the\eqcount#2})}}
%%%%%%%%%%%%%%%%%%%%%%%%%%%%%%%%%%%%%%%%%%%%%%%%%%%%%%%%%%%%%%%%%
%                                                               %
%       Hyperrefs:                                        	%
%                                                               %
%%%%%%%%%%%%%%%%%%%%%%%%%%%%%%%%%%%%%%%%%%%%%%%%%%%%%%%%%%%%%%%%%
\newtoks\url
\def\Href#1#2{\catcode`\#=12\url={#1}\catcode`\#=\active#2}
\def\href#1#2{{#2}}

%%%%%%%%%%%%%%%%%%%%%%%%%%%%%%%%%%%%%%%%%%%%%%%%%%%%%%%%%%%%%%%%%
%                                                               %
%       FORMAT:                                                 %
%                                                               %
%%%%%%%%%%%%%%%%%%%%%%%%%%%%%%%%%%%%%%%%%%%%%%%%%%%%%%%%%%%%%%%%%
\parskip=3.5pt plus .3pt minus .3pt
\baselineskip=14pt plus .1pt minus .05pt
\lineskip=.5pt plus .05pt minus .05pt
\lineskiplimit=.5pt
\abovedisplayskip=18pt plus 4pt minus 2pt
\belowdisplayskip=\abovedisplayskip
\hsize=14cm
\vsize=20cm
\hoffset=1.5cm
\voffset=1.8cm
\frenchspacing
\nopagenumbers
%%%%%%%%%%%%%%%%%%%%%%%%%%%%%%%%%%%%%%%%%%%%%%%%%%%%%%%%%%%%%%%%%
%                                                               %
%       VARIOUS DEFINITIONS                                     %
%                                                               %
%%%%%%%%%%%%%%%%%%%%%%%%%%%%%%%%%%%%%%%%%%%%%%%%%%%%%%%%%%%%%%%%%

\def\*{\partial}
\def\punkt{\,\,.}
\def\komma{\,\,,}

\def\+{\!+\!}
\def\={\!=\!}
\def\small#1{{\hbox{$#1$}}}
\def\half{\small{1\over2}}
\def\fraction#1{\small{1\over#1}}
\def\tr{\hbox{\rm tr}}
\def\eg{{\tenit e.g.}}
\def\ie{{\tenit i.e.}}

\def\id{1\hskip-3.5pt 1}

\def\nl{\hfill\break\indent}
\def\nlni{\hfill\break}

\def\\{\cr}
\def\*{\partial}
\def\a{\alpha}
\def\b{\beta}

\def\d{\delta}
\def\e{\varepsilon}

\def\g{\gamma}

\def\l{\lambda}
\def\m{\mu}
\def\n{\nu}
\def\r{\rho}
\def\s{\sigma}

\def\ki{\chi}
\def\D{\Delta}
\def\G{\Gamma}

\def\Z{{\Bbb Z}}
\def\R{{\Bbb R}}
\def\punkt{\,\,.}
\def\komma{\,\,,}

\def\II{\hbox{I\hskip-0.6pt I}}
\def\id{1\hskip-3.4pt 1}
\def\wdg{\!\wedge\!}
\def\fraction#1{\hbox{${1\over#1}$}}

\def\tr{\hbox{\rm tr}}
\def\eg{{\tenit e.g.}}
\def\ie{{\tenit i.e.}}

\def\eq#1{$$#1\eqn$$}
\def\eql#1#2{$$#1\Eqn#2$$}
\def\eqa#1{$$\eqalign{#1}\eqn$$}
\def\eqal#1#2{$$\eqalign{#1}\Eqn#2$$}

\def\ra{\rightarrow}

\headtext={Cederwall, Gran, Holm, Nilsson: ``Finite Tensor
Deformations$\ldots$''}

%%%%%%%%%%%%%%%%%%%%%%%%%%%%%%%%%%%%%%%%%%%%%%%%%%%%%%%%%%%%
%
%       THE PAPER
%
%
%%%%%%%%%%%%%%%%%%%%%%%%%%%%%%%%%%%%%%%%%%%%%%%%%%%%%%%%%%%%

\null\vskip-1cm
\hbox to\hsize{\hfill G\"oteborg-ITP-{\old98}-{\old19}}
\hbox to\hsize{\hfill\tt hep-th/9812144}
\hbox to\hsize{\hfill December, {\old1998}}

\vskip3cm
\centerline{\fourteencp Finite Tensor Deformations of Supergravity Solitons}
\vskip4pt
\vskip\parskip
\centerline{\twelvecp}

\vskip1.2cm
\centerline{\twelverm Martin Cederwall, Ulf Gran, Magnus Holm}
\centerline{\twelverm and Bengt E.W. Nilsson}

\vskip.8cm
\centerline{\it Institute for Theoretical Physics}
\centerline{\it G\"oteborg University and Chalmers University of Technology }
\centerline{\it S-412 96 G\"oteborg, Sweden}

\vskip.8cm
\catcode`\@=11
\centerline{\tentt 
	martin.cederwall,gran,holm,bengt.nilsson@fy.chalmers.se}
\catcode`\@=\active

\vskip2.2cm

\centerline{\bf Abstract}

{\narrower\noindent 
We consider brane solutions where the tensor degrees of freedom 
are excited. Explicit solutions to the full non-linear
supergravity equations of motion are given for the M{\old5} and D{\old3} 
branes, corresponding to finite selfdual tensor or Born--Infeld 
field strengths. The solutions are BPS-saturated and half-supersymmetric.
The resulting metric space-times are analysed.
\smallskip}
\vfill\eject

\ref\Goldstone{T. Adawi, M. Cederwall, U. Gran, B.E.W. Nilsson 
	and B. Razaznejad,\nl
	{\xit ``Goldstone tensor modes''}, \hepth{9811145}.}
\ref\KappaBranes{M.~Cederwall, A.~von~Gussich, B.E.W.~Nilsson 
	and A.~Westerberg,\nl
	{\xit ``The Dirichlet super-three-brane in type \II B supergravity''},
	\nl \NPB{490}{1997}{163} [\hepth{9610148}];\nlni
	M. Cederwall, A. von Gussich, B.E.W. Nilsson, P. Sundell
        and A. Westerberg, \nl{\xit ``The Dirichlet super-p-branes in
        type \II A and \II B supergravity''}, 
	\nl\NPB{490}{1997}{179} [\hepth{9611159}];\nlni
	M. Aganagic, C. Popescu and J.H. Schwarz, 
	{\xit ``D-brane actions with local kappa symmetry''}, 
	\nl\PLB{393}{1997}{311} [\hepth{9610249}];\nl
        {\xit ``Gauge-invariant and gauge-fixed D-brane actions''},
        \NPB{495}{1997}{99} [\hepth{9612080}];\nlni
	E. Bergshoeff and P.K. Townsend, {\xit ``Super D-branes''}, 
	\NPB{490}{1997}{145} [\hepth{9611173}].}
\ref\CW{M. Cederwall and A. Westerberg,
	{\xit ``World-volume fields, {\xrm SL(2;{\eightbbb Z})} and duality: 
		the type \II B 3-brane''},
	\nl\jhep{02}{1998}{004} [\hepth{9710007}].} 
\ref\CNS{M. Cederwall, B.E.W. Nilsson and P. Sundell, 
	{\xit ``An action for the 5-brane in D=11 supergravity''},
	\nl\jhep{04}{1998}{007} [\hepth{9712059}].} 
\ref\CHSKM{C.G. Callan, Jr., J.A. Harvey and A. Strominger,
	{\xit ``Worldbrane actions for string solitons''}, 
	\nl\NPB{367}{1991}{60};\nlni
	D.M. Kaplan and J. Michelson, {\xit ``Zero modes for the D=11 
	membrane and five-brane''}, 
	\nl\PRD{53}{1996}{3474} [\hepth{9510053}].}
\ref\ElevenSG{E. Cremmer, B. Julia and  J. Scherk,
	{\xit ``Supergravity theory in eleven-dimensions''},
	\nl\PLB{76}{1978}{409};\nlni
	L.~Brink and P.~Howe, {\xit ``Eleven-dimensional supergravity 
	on the mass-shell in superspace''},
	\nl\PLB{91}{1980}{384};\nlni
	E.~Cremmer and S.~Ferrara, 
	{\xit ``Formulation of eleven-dimensional supergravity 
	in superspace''},\nl\PLB{91}{1980}{61}.}
\ref\Duality{E.~Witten, 
	{\xit ``String theory dynamics in various dimensions''},
	\nl\NPB{443}{1995}{85} [\hepth{9503124}];\nlni
	C.M. Hull and P.K. Townsend,
        {\xit ``Unity of superstring dualities''},
        \nl\NPB{438}{1995}{109} [\hepth{9410167}];\nlni
	J.H. Schwarz, {\xit ``The power of M theory''},
        \PLB{367}{1996}{97} [\hepth{9510086}];\nlni
	A.~Sen, {\xit ``Unification of string dualities''},
	Nucl. Phys. Proc. Suppl. {\xbf58} ({\xold1997}) {\xold5}
	[\hepth{9609176}];\nlni
	P.K. Townsend, {\xit ``Four lectures on M-theory''},
        \hepth{9612121}.}
\ref\Guven{R.~G\"uven, 
	{\xit ``Black p-brane solutions of D=11 supergravity theory''}, 
	\PLB{276}{1992}{49}.}
\ref\Fivebranes{E. Witten, {\xit ``Five-brane effective action in M-theory''},
	\JGP{22}{1997}{103} [\hepth{9610234}];\nlni
	M. Aganagic, J. Park, C. Popescu and J.H. Schwarz,\nl
	{\xit ``World-volume action of the M theory five-brane''},
	\NPB{496}{1997}{191} [\hepth{9701166}];\nlni
	P.S.~Howe and E.~Sezgin, {\xit ``D=11, p=5''},
        \PLB{394}{1997}{62} [\hepth{9611008}];\nlni
        P.S.~Howe, E.~Sezgin and P.C.~West,\nl
        {\xit ``Covariant field equations of the M theory five-brane''},
        \PLB{399}{1997}{49} [\hepth{9702008}];\nl
        {\xit ``The six-dimensional self-dual tensor''},
        \PLB{400}{1997}{255} [\hepth{9702111}];\nlni
	E. Bergshoeff, M. de Roo and T. Ortin,      
        {\xit ``The eleven-dimensional five-brane''},\nl
        \PLB{386}{1996}{85} [\hepth{9606118}];\nlni
	I. Bandos, K. Lechner, A. Nurmagambetov, P. Pasti, D. Sorokin
        and M. Tonin,\nl {\xit ``Covariant action for the super-five-brane
                of M-theory''},
        \nl\PRL{78}{1997}{4332} [\hepth{9701037}];\nlni
	T. Adawi, M. Cederwall, U. Gran, M. Holm and B.E.W. Nilsson,\nl
	{\xit ``Superembeddings, non-linear supersymmetry and 5-branes''},\nl
	\IJMP{A13}{1998}{4691} [\hepth{9711203}].}
\ref\StringSolitons{M.J. Duff, R.R. Khuri and J.X. Lu,
	{\xit ``String solitons''},
	Phys. Report. {\xbf259} ({\xold1995}) {\xold213}  [\hepth{9412184}].}
\ref\HoweWest{P.S. Howe and P.C. West,
	{\xit ``The complete N=2, d=10 supergravity''},
	\NPB{238}{1984}{181}.}
\ref\DuffLu{M.J. Duff and J.X. Lu, {\xit ``The self-dual type \II B
	superthreebrane''}, \PLB{273}{1991}{409}.}
\ref\DuffReview{M.J. Duff, {\xit ``Supermembranes''}, \hepth{9611203}.}
\ref\BKOP{E. Bergshoeff, R. Kallosh, T. Ortin and  G. Papadopoulos,\nl
	{\xit ``Kappa-symmetry, supersymmetry and intersecting branes''},\nl
	\NPB{502}{1997}{149} [\hepth{9705040}].}
\ref\SorokinTownsend{D. Sorokin and P.K. Townsend,
    {\xit ``M-theory superalgebra from the M-5-brane''},
    \nl\PLB{412}{1997}{265} [\hepth{9708003}].}
\ref\StringSoliton{P.S. Howe, N.D. Lambert and P.C. West,
    {\xit ``The self-dual string soliton''},
    \nl\NPB{515}{1998}{203} [\hepth{9709014}];\nlni
	 J.P. Gauntlett, N.D. Lambert and P.C. West,
	{\xit ``Supersymmetric fivebrane solitons''},
	\hepth{9811024}.}
\ref\Holm{M. Holm, G\"oteborg-ITP-98-21, to appear on {\xtt hep-th}.}

%%%%%%%%%%%%%%%%%%%%%%%%BEGINNING OF TEXT%%%%%%%%%%%%%%%%%%%%%%%%

\section\intro{Introduction}The way in which branes in M theory and
string theory arise as ``soliton'' solutions of {\old11}- or 
{\old10}-dimensional supergravity is well known, see \eg\
[\StringSolitons,\DuffReview].
Much less explored
is the exact relation between the dynamics of the brane
degrees of freedom and the target space fields. The former of course
arise as zero-modes of the latter around a solitonic solution 
[\CHSKM,\Goldstone], but
when one goes beyond a linear approximation, no such relation has
been established so far. Part of the motivation of the present work is to
fill this gap. Specifically, we address the question of finding solutions
to the supergravity equations of motion corresponding to finite 
excitations of the tensorial degrees of freedom, while keeping the
brane flat and infinite. The analysis is applied to the M{\old5} brane
of {\old11}-dimensional supergravity and the D{\old3} brane of type \II B
supergravity, which both are truly solitonic. There are {\it a priori}
strong reasons to believe that analytic solutions exist, since they are
related to the dynamics of Born--Infeld vector fields and selfdual tensors
on the world-volumes of the D{\old3} and M{\old5} branes, respectively.
This calculation is carried through in section {\old2}. Section {\old3}
examines the metric properties of the resulting space-time, especially
a limiting case for maximal field strength, where no asymptotic Minkowski
region exists. In section {\old4}, we show that the solutions are
half-supersymmetric and construct the corresponding Killing spinors.

\section\zm{Finite tensor deformations}We want to find exact 
solutions for the M{\old5} and D{\old3} branes, where we have
finite field strength deformations. What makes it possible to find exact
solutions are the nice algebraic properties of the selfdual field strengths we
are dealing with.   
For most of our conventions and notation we refer to ref. [\Goldstone].
Here we just state our notation for the different types of indices occurring:
\hfill\break
Space-time indices: $M,N,\ldots$ (coordinate-frame), 
$A,B,\ldots$ (inertial);\hfill\break
Longitudinal indices: $\mu,\nu,\ldots$ (coordinate-frame), 
$i,j,\ldots$ (inertial);
\hfill\break
Transverse indices: $p,q,\ldots$ (coordinate-frame), 
$p',q',\ldots$ (inertial). 

\subsection\Mfive{The M5 brane}The {\old4}-form field strength $H$ 
should be parametrised by a closed {\old3}-form
$F(x)$ lying in the longitudinal directions, according 
to experience from brane dynamics [\Fivebranes,\CNS]. This can also be understood
from the general Goldstone analysis
[\Goldstone]. In contrast to the (infinitesimal) Goldstone analysis, where $F$
fulfilled a linear selfduality relation, $F$ should in the exact analysis
fulfill some non-linear selfduality relation. We are going to treat the
simplest case where $F$ is constant. 
Consider the equation of motion for $H$, $d{\star}H-{1\over2}H\wdg H=0$ [\ElevenSG].
The $\star$ operation involves the dualisation with the metric restricted
to the {\old6}-dimensional longitudinal directions.
In the Goldstone analysis [\Goldstone], where we considered
an infinitesimal excitation of $F$, this
metric was proportional to $\id$ and we did not have to care much about
whether we had the radius-independent tensor in coordinate-frame or inertial
indices, they just differed by a scalar function of the radial coordinate.
Now, the dualisation in coordinate-frame indices involves a metric that will
be ``non-trivial'', and for the selfduality to be consistent with 
radius-independence it must be possible to formulate it in terms of 
an inertial tensor.

Take $h_{ijk}$ to be a (linearly) anti-selfdual inertial tensor. Define
$q_{ij}={1\over2}h_i{}^{kl}h_{jkl}$. Then, $\tr\,q=0$ and $q^2=\mu\id$, where
$\mu={1\over6}\tr\,q^2$. The tensor $(qh)_{ijk}\equiv q_i{}^lh_{ljk}$ is
automatically antisymmetric and selfdual. For later purposes, we define
$\nu={1\over2}\sqrt\mu$. 
The most general Ansatz for the deformed {\old4}-form is now
$$
\eqalign{
H_{\mu\nu\l p}&=e_\mu{}^ie_\nu{}^je_\l{}^k\*_p\D F_{ijk}\komma\cr
F_{ijk}&=fh_{ijk}+g(qh)_{ijk}\komma\cr}
\Eqn\MAns
$$
where $f$ and $g$ are functions of $\mu$ and of the radial coordinate $\rho$. 
Due to the algebraic properties of $h$ all higher order terms can be 
reduced to the two terms in the Ansatz. The necessity to include the 
second term is that the radial 
derivative on ${\star}H$ acts not only on the tensor but also on the vielbeins.
The field along the {\old4}-sphere will not change, since the magnetic
charge should not be altered, so the background solution [\Guven] 
remains unaltered
\eq{
H_{pqrs}=\d^{tu}\e_{pqrst}\*_u\D\komma
}
where $\D$ is a harmonic function of the transverse coordinates, \ie, 
$\d^{pq}\*_p\*_q\D=0$. 
By considering all functions, as $f$ and $g$ above,
as functions of $\D$ instead of $\r$, one covers AdS space 
($\D=({R\over\r})^3$) as well as the asymptotically
flat brane solutions ($\D=1+({R\over\r})^3$), without any extra 
calculational complication.

As an Ansatz for the vielbeins, we take
\eqa{&e_\mu{}^i=\d_\mu{}^j(a\d_j{}^i+bq_j{}^i)\komma\cr
	&e_p{}^{p'}=c\d_p{}^{p'}\komma\cr
}
where $a$, $b$ and $c$ are functions of $\mu$ and $\D$. One thing that makes 
the calculations simpler is that all matrices that may occur, vielbeins
and derivatives of vielbeins, commute with each other. 
One may quite easily calculate the Ricci tensor. A first observation is
that the RHS of Einstein's equations can never contain $\*_p\*_q\D$, so
such terms must not be present in $R_{pq}$. This implies
that $c=(\det e_\mu{}^i)^{-1/\tilde d}$, where 
$\tilde d=D-d-2$\foot\star{{\oldsize D} is the target space dimension and 
{\xold d} that of the brane. Thus, in this case $\tilde{\hbox{\xold d}}$=3.}. 
When this is used, the Ricci tensor
is, expressed in terms of $A\equiv\log e$ ($e$ denoting $e_\mu{}^i$),
\eqal{	R_{pq}&=-\*_p\D\*_q\D\left(\tr(A'{}^2)
	+\fraction{\tilde d}(\tr A')^2\right)
		+\fraction{\tilde d}\d_{pq}(\*\D)^2\tr A''\komma\cr
	R_{\mu\nu}&=-c^{-2}(\*\D)^2e_\mu{}^ie_\nu{}^jA''_{ij}\punkt\cr
}{\Ricci}
Prime denotes differentiation w.r.t. $\D$ and $(\*\D)^2\equiv\d^{pq}\*_p\D\*_q\D$.
The matrix $A$ will be parametrised as $A={1\over d}(\a\id+\b q)$, and $\a$ is
actually equal to $\log\det e_\mu{}^i$.
It is also convenient to rescale the functions in the Ansatz for $H$ as
$\phi=e^{-\a}f$, $\psi=e^{-\a}g$.
The remaining part of Einstein's equations, together with the e.o.m.
for $H$, are now
\eqa{0&=\a''-e^{2\a}(1-2\mu\phi\psi)	\komma\cr
	0&=\b''+3e^{2\a}(\phi^2+\mu\psi^2)	\komma\cr
	0&=\a'^2+\fraction3\mu\b'^2-e^{2\a}(1-4\mu\phi\psi)	\komma\cr
	0&=\phi'+(e^\a+\half\a')\phi-\half\mu\b'\psi	\komma\cr
	0&=\psi'-(e^\a-\half\a')\psi-\half\b'\phi	\punkt\cr
}
This is one equation too many, but by differentiating the third
equation one gets a combination of the other four (eventually, one has
to check that the integration constant vanishes). 
The $\mu$-dependence can be removed by redefining
$\mu^{1/4}\phi\ra\phi$, $\mu^{3/4}\psi\ra\psi$, $\mu^{1/2}\b\ra\b$;
the equations become identical to the ones above with $\mu=1$.

The background solution, describing either AdS${}_7\times S^4$  
or an M{\old5} brane with no tensor excitations, is $\a=-\log\D$ and the rest zero.
If one builds up the solution order by order in the perturbation, one first solves the
zero-mode equation for $\phi$ giving $\phi=k\D^{-1/2}$. This linearised
solution then backreacts on the geometry giving the lowest order
perturbation to $\b\sim\D^{-1}$. This non-diagonal metric then forces
the tensor to contain the other duality component, $\psi\sim\D^{-3/2}$,
which in turn enforces a diagonal modification to the vielbein, \ie\ of $\a$, of
the order $\D^{-2}$. And so it goes on. This becomes an expansion in
negative powers of $\D$ and at the same time in the constant $k$, 
which just determines
the normalisation of $h_{ijk}$. The $\mu$-dependence is reinserted 
by choosing $\mu^{-1/4}k=1$ (so that $\phi$ starts out with
$\D^{-1/2}$), which makes the expansion look like
\eqa{
	&\a\sim-\log\D+\mu\D^{-2}+\mu^2\D^{-4}+\ldots\cr
	&\b\sim\D^{-1}+\mu\D^{-3}+\mu^2\D^{-5}+\ldots\cr
	&\phi\sim\D^{-1/2}+\mu\D^{-5/2}+\mu^2\D^{-9/2}+\ldots\cr
	&\psi\sim\D^{-3/2}+\mu\D^{-7/2}+\mu^2\D^{-11/2}+\ldots\cr
}
Considering the first few terms in this expansion enabled us 
to find the exact solution:
\eqa{\a&=-{1\over2}\log(\D^2-\nu^2)\komma\cr
\b&={3\over4\nu}\log{\D-\nu\over\D+\nu}\komma\cr
\phi&={1\over2}\Bigl({1\over\sqrt{\D+\nu}}+{1\over\sqrt{\D-\nu}}\Bigr)\komma\cr
\psi&={1\over4\nu}\Bigl({1\over\sqrt{\D+\nu}}-{1\over\sqrt{\D-\nu}}\Bigr)
\punkt\cr
}

Before inserting the explicit solution for $\a$ and $\b$ in the metric, it
is useful to note that the eigenvalues of the matrix $q$ are $\pm2\nu$, and
that there are three of each. 
We group the longitudinal coordinates 
accordingly into $x_\pm$. The time direction is included in $x_-$.
The metric then becomes
$$
ds^2=(\D^2-\nu^2)^{-1/6}\left[\left(\D+\nu\over\D-\nu\right)^{1/2}dx_-^2
	+\left(\D-\nu\over\D+\nu\right)^{1/2}dx_+^2\right]
	+(\D^2-\nu^2)^{1/3}dy^2\punkt\Eqn\ExplicitFiveMetric
$$
It clearly reduces to the well known M{\old5} brane metric 
when the tensor deformation
is absent, \ie, when $\nu=0$. We will return to the properties of the 
metric in section {\old3}. Finally, inserting the solution into 
the Ansatz (\MAns) gives us the {\old4}-form in inertial indices:
\eq{	H_{ijkp'}={\d_{p'}{}^{p}\*_p\D\over(\D^2-\nu^2)^{2/3}}
	\left[ {1\over\sqrt{\D+\nu}}\Pi_+h
	 + {1\over\sqrt{\D-\nu}}\Pi_-h\right]_{ijk}\komma
}
where $\Pi_\pm={1\over2}(\id\pm{q\over2\nu})$ project all indices
on the $+$ and $-$
directions (the algebraic properties of $h$ tell us that only
$h_{i_+j_+k_+}$ and $h_{i_-j_-k_-}$ are non-vanishing).

\subsection\Dthree{The D3 brane}The relevant tensor field in type \II B
supergravity [\HoweWest] is the complex {\old3}-form field strength $H$. 
The D{\old3} brane is
invariant under SL(2;$\Z$) transformations, and it is convenient to keep
SL(2;$\Z$) covariance throughout the calculations. The Bianchi identity 
and equation of motion for $H$ are
$$
\eqalign{
	&DH-P\wdg\bar H=0\komma\cr
	&D{\star}H-P\wdg{\star}\bar{H}+iG\wdg H=0\komma\cr
}
\Eqn\HBianchiEOM
$$
where the U(1) covariant derivative $D$ contains a connection $Q$, which
together with $P$ are the left-invariant SL(2;$\R$) 
Maurer--Cartan forms built from the scalars\foot\star{We will 
leave {\xold Q} out of the
continued discussion---to the initiated reader it will be obvious that
it is pure gauge, and we use this to put it to zero.}. 

We use an Ansatz analogous to the M{\old5} brane case:
\eql{H=d\D\wdg\tilde F\komma
}{\DAns}
where
\eq{\tilde F_{ij}=fF_{ij}+g\bar{F}_{ij}\punkt
}
We again have one anti-selfdual ($\star F=-iF$) and one selfdual 
($\star\bar F=i\bar F$) part. 
The algebraic properties of the matrix $F$ are
\eqa{	(FF)_{ij}&=\mu\d_{ij}\,\,;\quad\mu=\fraction4\tr F^2\komma\cr
	(F\bar{F})_{ij}&=(F\bar{F})_{ji}\komma\cr
	\tr(F\bar{F})&=0\punkt
} 

When we excite $H$ we must also excite the {\old1}-form $P$, 
as a consequence of the equations of
motion, and we need an Ansatz for that too,
\eql{P=ud\D\komma
}{\PAns}
where $u=u(\mu,\bar{\mu},\D)$. The Bianchi identity and equation of motion
for $P$ are
$$
\eqalign{
	&DP=0\komma\cr
	&D{\star}P-H\wdg{\star}H=0\punkt\cr
}
\Eqn\PBianchiEOM
$$
The Ans\"atze trivially fulfill the Bianchi identity parts of (\HBianchiEOM)
and (\PBianchiEOM) when only functions of the radial coordinate are 
considered.

The Ansatz for the vielbeins are
\eqa{&e_\mu{}^i=\d_\mu{}^j(a\d_j{}^i+b(F\bar{F})_j{}^i)\komma\cr
	&e_p{}^{p'}=c\d_p{}^{p'}\komma
}
which is also completely analogous to the M{\old5} brane case, 
\ie, $e_\mu{}^i$ is made
up of the two symmetric matrices we can construct.

The Ricci tensor is given by eq. (\Ricci), where A is now parametrised as
$A=\fraction{d}(\a\id+\b F\bar{F})$. The equations of motion we want to
solve are Einstein's equation
$$
\eqalign{
	R_{MN}&=2\bar{P}_{(M}P_{N)}+\bar{H}_{(M}{}^{RS}H_{N)RS}-
		\fraction{12}g_{MN}\bar{H}_{RST}H^{RST}\cr
		&+\fraction{96}G_{(M}{}^{RSTU}G_{N)RSTU}\komma\cr
}
\eqn
$$
together with the the equations of motion in (\HBianchiEOM) and (\PBianchiEOM).
We use the background solution [\DuffLu]
\eqa{	G&=\pm\fraction{5!}(\d^{mn}\*_m\D\e_{npqrst}dy^p\wdg
	dy^q\wdg dy^r\wdg dy^s\wdg dy^t\\
	&\quad\thinspace-5g^{-2}\*_m\D\e_{\m\n\r\s}dy^m\wdg dx^{\m}\wdg
	dx^{\n}\wdg dx^{\r}\wdg dx^{\s})\komma\\
}
where $g=\det(g_{MN})$ (the first term, which gives the D{\old3} brane charge,
is identical to the one in the ordinary D{\old3} brane solution, and the second
is its dual, where we have taken into account that the metric is modified).
With the same rescalings as for the M{\old5} brane, \ie, $f=e^\a
\phi$, $g=e^\a \psi$ and $u=e^\a \ki$, we can rewrite the equations 
of motion as
\eqa{	0&=\a''-e^{2\a}\Bigl(1+4(\mu\phi\bar{\psi}
		+\bar{\mu}\bar{\phi}\psi)\Bigr)	\komma\cr
	0&=\b''-8e^{2\a}(\phi\bar{\phi}+\psi\bar{\psi})	\komma\cr
	0&=\a'^2+\fraction2\mu\bar{\mu}\b'^2-e^{2\a}
		\Bigl(1+8(\mu\phi\bar{\psi}+\bar{\mu}\bar{\phi}\psi)
		-4\ki\bar{\ki}\Bigr)	\komma\cr
	0&=\phi'+(e^\a+\half\a')\phi-\half\bar{\mu}\b'\psi+e^\a
		 \ki\bar{\psi}	\komma\cr
	0&=\psi'-(e^\a-\half\a')\psi-\half\mu\b'\phi+e^\a
		 \ki\bar{\phi}	\komma\cr
	0&=\ki'+\a'\ki+2e^\a(\mu\phi^2+\bar{\mu}\psi^2)	\punkt\cr}
By differentiating the third equation we get a combination of the other five.
The first three equations come from Einstein's equation, the fourth and fifth
from the equation for $H$ and the last one from the equation for $P$.
From the properties of the fields involved under U(1) gauge transformations
it is clear that $\a$, $\b$ and $\phi$ are real functions, while
$\psi$ and $\ki$ must be real functions multiplied by $\mu$.
The solution to the equations is given by
\eqa{
\a&=-{1\over2}\log(\D^2-\nu^2)\komma\cr
\b&=-{2\over\nu}\log{\D-\nu\over\D+\nu}\komma\cr
\phi&={1\over2}\Bigl({1\over\sqrt{\D+\nu}}+{1\over\sqrt{\D-\nu}}\Bigr)\komma\cr
\psi&=-{\mu\over\nu}\Bigl({1\over\sqrt{\D+\nu}}
	-{1\over\sqrt{\D-\nu}}\Bigr)\komma\cr
\ki&={\mu\over\sqrt{\D^2-\nu^2\komma}}
}
where $\nu=2|\mu|$ and we have used the normalisation that 
$\phi \ra \D^{-1/2}$ as $\mu
\ra 0$ (the same rescaling argument holds here as for the M{\old5} brane).

The metric may be diagonalised in the same manner as the M{\old5} brane metric
(the eigenvalues of $F\bar F$ are $\pm{\nu\over2}$, and now time is
in the positive eigenvalue sector), giving
$$
ds^2=(\D^2-\nu^2)^{-1/4}\left[\left(\D+\nu\over\D-\nu\right)^{1/2}dx_+^2
	+\left(\D-\nu\over\D+\nu\right)^{1/2}dx_-^2\right]
	+(\D^2-\nu^2)^{1/4}dy^2\punkt\Eqn\ExplicitThreeMetric
$$
Inserting the solution into the Ans\"atze (\DAns) and (\PAns) finally gives us
\eqa{	H_{p'ij}={\d_{p'}{}^{p}\*_p\D\over2(\D^2-\nu^2)^{5/8}}&
	\left[ {1\over\sqrt{\D+\nu}}(F-{2\mu\over\nu}\bar{F})
	 + {1\over\sqrt{\D-\nu}}(F+{2\mu\over\nu}\bar{F}) \right]_{ij}\komma\cr\cr
	&P_{p'}={\mu\d_{p'}{}^{p}\*_p\D\over(\D^2-\nu^2)^{9/8}}
}
in inertial indices.
 
We want to stress that the structures of the solutions for the 
D{\old3} and M{\old5}
branes are completely analogous (except that we happen to excite additional
scalar fields in the D{\old3} brane case, which however is easily dealt with).
The linear terms in the deformations, \ie, the lowest order terms
in the series expansions of $\phi$, agree with the zero-modes derived
in ref. [\Goldstone].

\section\Metrics{Properties of the metrics}The metric space-times
described by eqns. (\ExplicitFiveMetric) and (\ExplicitThreeMetric) 
represent deformations of the original
AdS$\times$sphere or brane space-times parametrised by one real
number $\nu$, measuring the square of the field strength. When the radial
coordinate $\r$ runs from 0 (which is the horizon in the brane case
and a subset of no special significance in the AdS case) to $\infty$,
$\D$ runs from $\infty$ to 1 for the brane and from $\infty$ to zero for AdS.
We see that there is potential danger when $\Delta-\nu$ becomes negative.

Let us first treat the AdS case. Here $\D-\nu=({R\over\r})^{\tilde d}-\nu$, and
this is bound to change sign at some finite radius when $\nu>0$. 
The question is
whether this is a physical singularity or not. It is straightforward to
calculate \eg\ the curvature scalar, and find that it diverges at this
radius. Such solutions do not define sensible space-times.

For the brane solutions, $\D-\nu=({R\over\r})^{\tilde d}+1-\nu$.
The solution makes sense for $\nu\leq1$. This is a reflection of the
Born--Infeld or Born--Infeld-like dynamics, which breaks down at
field strengths where $\det(g+F)$ vanishes. The behaviour of the
solutions for small radii is always unmodified, \ie,
AdS${}_{d+1}\times S^{\tilde d+1}$. For large radii, there is an
asymptotic Minkowski region as long as $\nu$ is strictly smaller than 1.

The limiting case, $\nu=1$, has some interesting properties.
One may calculate the curvature scalar, and find that it is non-singular
as $\r\ra\infty$; it goes asymptotically as $\r^{-1}$. After some
trivial rescalings, the leading terms in the metric behave as 
$$
\eqalign{
\hbox{M{\old5}:}\qquad ds^2&=\r^2dx_-^2+\r^{-1}(dx_+^2+dy^2)\komma\cr
\hbox{D{\old3}:}\qquad ds^2&=\r^3dx_+^2+\r^{-1}(dx_-^2+dy^2)\punkt\cr
}
\eqn
$$
As $\r\ra\infty$, half of the longitudinal directions ``expand'' and
the other half ``shrink'', and what remains is something rather
like a continuously smeared membrane or string, respectively. Whether
this interpretation is physically relevant is unclear to us, however
it is supported by the asymptotic behaviour of the dual of the 
tensor field, which asymptotically lies in the shrinking directions and
the $(\tilde d+1)$-sphere.
The limiting metric does not factorise, but it has some things in common
with the AdS metric: the space-like distance to $\r=\infty$ is infinite,
but light rays may reach infinity (and come back) in finite time.

\section\Killing{Supersymmetric properties of the solutions}In the absence
of an expectation value for the field strength on the brane, it is well
known that the solutions break half the supersymmetry, \ie, that there
are {\old16} Killing spinors. Arguing na\"\i vely in terms of the
field theory on the brane, one might expect that
giving a background value to $F$ would break the entire remaining 
global supersymmetry, so that the solutions presented here would be
non-supersymmetric (and perhaps less interesting). What actually happens
is instead that there are new combinations of the broken and unbroken
supersymmetries that become Killing spinors in the presence of $F\neq0$,
and that the new solutions enjoy the same amount of supersymmetry,
{\old16} Killing spinors. 

There are at least two good arguments why this
should happen. The first, more conceptual, is that the tensor modes
are very much on the same footing as the scalar ones, in the sense
that they all result from breaking of large gauge transformations 
[\Goldstone]. Deforming a brane by giving constant ``field strength''
to scalars (transverse coordinates) corresponds to tilting the brane
through some angle, a somewhat trivial operation that of course does
not change the number of supersymmetries. The definition of world-volume
chirality however changes, and one has to recombine broken and unbroken
supersymmetries to recover the new Killing spinors. A similar phenomenon
should occur
for the tensors, and we already know that an analogous mechanism is at
work for the tensors themselves, where chirality (selfduality) becomes
nonlinear. The second, more technical, argument is that one knows from
work on $\kappa$-symmetry in 
supersymmetric brane dynamics  [\KappaBranes,\Fivebranes,\CW,\CNS]
that there is a half-rank projection matrix,
or generalised chirality operator [\BKOP],
acting on spinors separating broken and unbroken supersymmetry, and that
this matrix generically depends on $F$. For constant $F$, this means that
there should be {\old16} global supersymmetries.

When the tensor degrees of freedom are turned on, the branes carry
not only magnetic charge, but also local electric charge
[\SorokinTownsend,\Goldstone].
The BPS property expressed through the existence of a local projection
on the Killing spinors involves both charges, which explains why the
excited brane may be BPS-saturated although the tensor excitations carry
energy. The configurations carrying global electric charge are 
world-volume solitons [\StringSoliton].

The most convincing argument is of course to construct the Killing
spinors explicitly, which we now proceed to do (although we satisfy ourselves
with the M{\old5} brane case).
The preserved supersymmetry obeys the Killing spinor equation obtained 
by setting the variation of the gravitino field in the background to zero:
$$
\d_\zeta{\psi}_{M}={D}_{M}{\zeta}
	-\fraction{288}({\G}_{M}{}^{NPQR}
	-8\d_M{}^N{\G}^{PQR}){\zeta}H_{NPQR}=0\punkt\eqn
$$
The inertial gamma matrices are split as $\G_A=(\g_i,\g_7\oplus\Sigma_{p'})$
The calculation is straightforward (along the lines of ref. [\Goldstone]).
After assuming that the only functional dependence comes through $\D$,
one obtains a differential equation for $\zeta$,
$$
\zeta'+\bigl[\fraction3\D(\D^2-\nu^2)^{-1}
	+\fraction4(\D^2-\nu^2)^{-1/2}\g_7\bigr]\zeta=0\komma
\Eqn\ZetaDiff
$$
and an algebraic condition 
$$
\half(\id+\G)\zeta=0\,;\quad
	\G=\D^{-1}(\D^2-\nu^2)^{1/2}
	\bigl(\g_7+\fraction{12}(\D^2-\nu^2)^{1/2}F_{ijk}\g^{ijk}\bigr)
\punkt\Eqn\SpinorProj
$$
It is now crucial that the last equation projects $\zeta$ on half the
original number of components. Using the explicit forms of the functions
entering into $F$ gives $\G^2=\id$, so that eq. (\SpinorProj) is a projection.
It defines a generalised chirality condition, which for any fixed radius
takes the form known from the $\kappa$-symmetric formulation of the M{\old5}
brane [\CNS]. The chirality condition varies continuously with the radial
coordinate, as does the non-linear selfduality condition on $F$.

The solution to eq. (\ZetaDiff) is
$$
\eqalign{
\zeta_-&=(\D^2-\nu^2)^{1/12}
\left({1\over\sqrt{\D+\nu}}+{1\over\sqrt{\D-\nu}}\right)^{1/2}\lambda_-
	\komma\cr
\zeta_+&=(\D^2-\nu^2)^{-5/12}
\left({1\over\sqrt{\D+\nu}}+{1\over\sqrt{\D-\nu}}\right)^{-1/2}\lambda_+
	\komma\cr
}
\Eqn\ZetaSol
$$
where $\zeta$ has been split in chirality components according to the
eigenvalue of $\g_7$ and where $\lambda_\pm$ do not depend on $\D$.
We notice that in the absence of a tensor field we recover the Killing
spinors of ref. [\Goldstone] which was $\zeta=\D^{-1/12}\lambda_-$.
It remains to be checked that the solutions (\ZetaSol) are consistent
with the chirality (\SpinorProj), \ie, that the $\D$-dependence cancels
upon inserting the solutions into the chirality condition. This indeed
happens, and the chirality condition condenses into
$$
\lambda_+=-\fraction{12}h_{ijk}\g^{ijk}\lambda_-\komma\eqn
$$
which together with eq. (\ZetaSol) gives the explicit form of the 
Killing spinors.

\section\sc{Discussion}We have derived a new class of 
half-supersymmetric solutions of 
{\old11}-dimensional and type \II B supergravity, corresponding to
M{\old5} and D{\old3} branes with non-vanishing constant field strength.
The structure of the solutions clearly reflects the property of 
Born--Infeld-like dynamics as opposed to quadratic actions, in that there
is a maximal allowed value of the field strength.

It is interesting to note that although the symmetry of the solutions
is smaller than in the case of vanishing field strength---the longitudinal
SO(1,5) part of the isometry group is broken into SO(1,2)$\times$SO(3) for the
M{\old5} brane (and accordingly for the D{\old3} brane), the amount of 
supersymmetry is unchanged (the longitudinal translations of course 
remain unbroken). The split of the longitudinal directions
in two groups is a novel property of brane solutions. It is 
not related to the longitudinal symmetry breaking induced by world-volume
solitons, rather this split seems to have something to do with 
other branes, in these cases membranes and strings. The phenomenon might
deserve further study, especially in the strong field limit. The
formalism of ref. [\Holm] may be useful in this context.

It should be possible to push the analysis further by considering also
configurations with field strengths that depend on the longitudinal
coordinates and thus derive the dynamics of the fields (the result would
be in the selfdual form of refs. [\CW,\CNS]).
Another application would be the generalisation to other types of
branes---the method presented here might provide a manifestly 
SL(2;$\Z$)-covariant formulation of the type \II B {\old5}-branes.
Finally, it would be interesting to understand whether the limiting
solutions of maximal field strength have some physical significance,
considering their interesting asymptotic structure.

%\acknowledgements 
%\vfill\eject
\refout

\end